# An Enrichment Method for Obtaining Biologically Significant Genes from Statistically Significant Differentially Expressed Genes in Comparative Transcriptomics

Panpaki Seekaki[1] and Norichika Ogata[1,2]

[1] Nihon BioData Corporation, 3-2-1 Sakado, Takatsu-ku, Kawasaki, Kanagawa, Japan
[2] Medical Mechanica Incorporated, 3-2-1 Sakado, Takatsu-ku, Kawasaki, Kanagawa, Japan
`norichik@nbiodata.com`

**Abstract.** Cells coordinate adjustments in genome expression to accommodate changes in their environment. A drug in culture media for *in vitro* preclinical testing sometimes cause drastic regime shifting of genome expression system depending on the concentrations; e.g. primary cultured cells exposed to high concentrations of phenobarbital (≥1.0 mM) recovered their tissue-specific character as part of an individual organism. Drastic changes of transcriptomes interrupt discovering biologically significant genes in comparative transcriptomics. Here, we compared the amount of environmental changes and the amount of transcriptome changes using phenobarbital and the Chinese hamster ovary derived established continuous cell line CHO-K1; immortalized cell lines are accepted for *in vitro* preclinical testing then primary cultured cells.

## 1   Introduction

Gene screenings are performed in comparative transcriptomics using RNA-seq. Statistically significant differentially expressed genes were obtained from every comparison between any transcriptomes. Some p values from a comparison of one replicate with another one must be below 0.05 in comparative transcriptomics with Type-I error control. Between replicates, no genes are truly differentially expressed, and the distribution of p-values is expected to be uniform in the interval [0,1] (Anders & Huber 2010). We had reviewed and accepted papers performing differentially gene expression analyses between samples, listing over hundreds of statistically significant differentially expressed genes and introducing studies dealing with functions of those genes. Someone performed meaning-less Gene Ontology analyses or pathway analyses except a few studies (Goecks et al. 2013). A few studies validated listed statistically significant differentially expressed genes (Lee et al. 2016), since most biologists may think there are few biologically significant genes in the statistically significant differentially expressed genes in comparative transcriptomics.

The problems of the comparisons in biology were discussed for a long time (Goethe 1820). Goethe pointed the problem of the comparisons in Osteology, osteologists may for the present state maintain that the division of the human bony structure was merely accidental; So that in descriptions, sooner or later, fewer bones were assumed, each describing them as they pleased and their own order. Finally, one wanted to deny man his "intermediate bones". He remarks that his osteological studies awakened in him the desire "to establish a type according to which all mammals would be able to be examined for their similarities and differences." (Gray 2004). We have to discover that for a meaning-full comparison, which are common to all subjects, and in which these parts are different. If we compared a human hand bone and a monkey's head bone, we would obtain many statistically significant differences and few biological significant differences. We have to compare the same bones of human and monkey and it is clear in Osteology, but not in comparative transcriptomics. A type of transcriptomes does not exist.

Previous studies indicated that the genome expression system displays a drastic regime shift (Luscombe et al. 2004; Noori 2014). Distinguish common and different is difficult in comparative transcriptomics. However, it was possible to judge transcriptomes into same regime or under different regime. Comparative transcriptomics for *in vitro* preclinical testing are widely performed as an alternative to animal tests. Determination of the drug concentration is critically important for *in vitro* drug-exposure testing in preclinical toxicology. High drug concentrations can induce a radical transcriptome response, while mild concentrations can make it difficult to determine cell responses. Ideally, a drug concentration should be high enough to induce the desired main effect, while not eliciting too many side effects. Previously, we had been forced to determine appropriate drug concentrations by considering cell morphology and through various other examinations.

A previous study established a quantitative method to determine the drug concentrations that should be used for *in vitro* preclinical testing using Shannon's information entropy of transcriptome data, judging transcriptomes into same regime or under different regime (Ogata et al. 2015). Shannon's information entropy a dimensionality reduction method was applied to obtain index of regime of the transcriptome. Dimensionality Reduction methods e.g. Principle component analysis and t-SNE had been used to transcriptome analyses, but biological meanings of value of results of these methods was unclear. Shannon's information entropy also reduces the dimensions of transcriptome data and the biological meaning is clear; a cellular dedifferentiation increases Shannon's information entropy of transcriptome data of the cells and the cellular differentiation decreases them (Ogata et al. 2012). The previous study discussed the relationship between the amount of transcriptome change and the amount of environmental change using silkworm fat body tissues cultured with several concentrations of phenobarbital and primary cultured cells exposed to high concentrations of drug recovered their tissue-specific character as part of an individual organism.

Previously, a perfect drug concentration was determined for primary cultured tissues. However, an industrial applicability of Shannon's information entropy is questionable, since established cell lines mainly accepted for *in vitro* preclinical testing.

Established cell lines partially lose their tissue-specific character as part of an individual organism; for example, transplanted HEK293T cell line derived pluripotent cells formed tumor-like structures contained ectoderm and mesoderm tissues in dorsal flanks of kidney in busulfan-treated mice, however, no typical endoderm structures (Lin et al. 2008; Wang et al. 2014). A follow-up study on regime shift using immortalized cell lines is needed.

Even from a theoretical point of view, there is room for argument the application of Shannon's information entropy on transcriptome analyses, since Shannon's information entropy ignore zero count data contained transcriptome data obtained using RNA-seq. Recent study estimated Kolmogorov complexity of transcriptome treating zero count data (Seekaki & Ogata 2017). The elementary theories of Shannon information and Kolmogorov complexity have a common purpose (Grunwald & Vitanyi 2010). Kolmogorov complexity is the minimum number of bits from which a particular message or file can effectively be reconstructed. Since zero count data contained transcriptome data may have some message, we should treat zero count data using Kolmogorov complexity. Similar transcriptomes that were not distinguished using Shannon's information entropy were distinguished using Kolmogorov complexity.

Here, we performed a comparison between the amount of transcriptome change and the amount of environmental change using an immortalized cell line CHO-K1 derived from Chines Hamster Ovary cells cultured with several concentrations of phenobarbital. We calculated Shannon's information entropy and Kolmogorov complexity of tranrciptome data. We also obtained phenobarbital-induced genes from a differentially expressed gene analysis.

## 2     Materials and Methods

All of the chemicals used in this study were of analytical grade. Phenobarbital sodium (Wako Pure Chemical, Osaka, Japan) was dissolved in distilled water to make a stock solution, which was added to the medium to make final concentrations of 0.25 and 2.5 mM phenobarbital. The original medium was replaced with phenobarbital-containing medium in the induction assays. CHO-K1 cells obtained from ATCC were incubated with 0, 0.25 and 2.5 mM phenobarbital for 10 hours. Induction assays were terminated by using QIAshredder (Qiagen, Basel, Switzerland), and the tissues were kept at −80°C until analysis. RNA was extracted using a commercial kit (RNeasy mini kit; Qiagen, Basel, Switzerland). We prepared a library for conventional RNA-seq using a commercial kit (TruSeq RNA Sample Kit; Illumina, San Diego, CA, USA) in accordance with the manufacturer's protocol. We sequenced libraries for conventional RNA-seq using a commercial sequencer (HiSeq 2000; Illumina) in accordance with the manufacturer's protocol. Short-read data have been deposited in the DNA Data Bank of Japan (DDBJ)'s Short Read Archive under project ID DRA005765. All raw sequencing reads were mapped to CHO-K1 RefSeq assembly (ID GCF_000223135.1) using bowtie (version 0.12.8) with the modified parameters (-l 75 -n 2 -p 4). Shannon's entropy of transcriptome data and Kolmogorov complexity were measured as previously described (Ogata et al. 2015; Seekaki & Ogata 2017). All of the data were processed using bash (version 3.2) and visualized using R. Linear regression analyses were performed using R and Fitting Linear Models (lm). Differentially expressed gene (DEG) analyses were performed using R and the TCC package (version 1.1.99) (Sun et al. 2013).

## 3     Results and Discussion

To investigate the effects of phenobarbital concentration on Shannon's information entropy and Kolmogorov complexity of trancriptome data, we sequenced 9 transcriptomes from CHO-K1 cells exposed to phenobarbital. Freshly subcultured cells ($2 \times 10^5$) were cultured for 24 hours in DMEM/F12 medium, and then cultured for 14 hours in medium supplemented with 0, 0.25 and 2.5 mM phenobarbital. We measured the Shannon's information entropy and Kolmogorov complexity of those transcriptomes. Shannon's information entropy were 11.99634, 11.96654, 11.88143 with 0 mM phenobarbital, 11.9747, 11.9967, 12.01303 with 0.25 mM phenobarbital and 12.04395, 12.01875, 12.06603 with 2.5 mM phenobarbital. The slope of the regression line is significantly different than 0 (p=0.0335). Values for the slope was 0.03138. The AIC of the model was -28.61618 (Figure 1a). Kolmogorov complexity were 0.0641145, 0.0641105, 0.0637179 with 0 mM phenobarbital, 0.0640257, 0.0642249, 0.0637159 with 0.25 mM phenobarbital and 0.0647754, 0.0645268, 0.0641362 with 2.5 mM phenobarbital. The slope of the regression line is significantly different than 0 (p=0.028). Values for the slope was $2.068e^{-04}$. The AIC of the model was -119.8417 (Figure 1b). The AIC of the model using Kolmogorov complexity was lower than that of the model using Shannon's information entropy. Kolmogo-rov complexity was better index of transcriptome data than Shannon's information entropy. Kolmogorov complexity changed between phenobarbital concentrations of 0.25 and 2.5 mM (Figure 1b). In previous study, cells cultured in media containing lower drug concentrations than the tipping point showed uniformly high Shannon's information entropy, while those cultured at higher drug concentrations than the tipping point showed uniformly low Shannon's information entropy. However, in this study, cells cultured in media containing low drug concentrations showed low Shannon's information entropy and those cultured at high drug concentrations showed low Shannon's information entropy. This contradiction indicating that there is unknown mechanisms in a drastic regime shift of the genome expression system. Traditional RNA-seq measures means of cultured cells and the changes of cellular heterogeneity also changes the RNA-seq results. There is a possibility that phenobarbital induction make increase cellular heterogeneity of immortalized cell lines. Single cell transcriptome analyses of cells displaying a drastic regime shift would discover this problem.

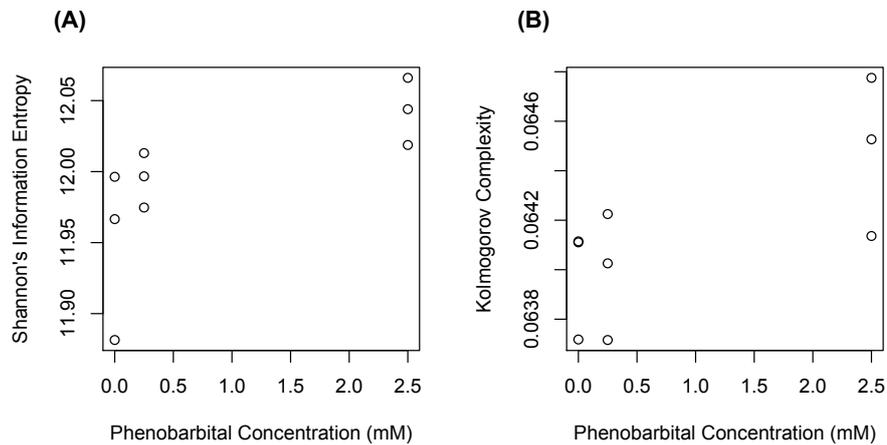

**Fig. 1. Comparisons between the amount of transcriptome change and the amount of environmental change.** (a) Scatter plot of drug concentration vs Shannon's information entropy. Transcriptomes of CHO-K1 cells that were cultured 10 hours in DMEM/F12 media supplemented with 0, 0.25 and 2.5 mM phenobarbital. (b) Scatter plot of drug concentration vs Kolmogorov complexity. Transcriptomes of CHO-K1 cells that were cultured 10 hours in DMEM/F12 media supplemented with 0, 0.25 and 2.5 mM phenobarbital.

We compared transcriptomes with close Kolmogorov complexity focusing on differentially expressed genes. In this study, a comparison between the transcriptome of a control group (0 mM phenobarbital) and that of an experimental group treated with the lowest phenobarbital concentration (0.25 mM) represented the comparison of close Kolmogorov complexity (Figure 2a). Indeed, only single differentially expressed gene was detected in the comparison between the control and the 0.25 mM phenobarbital-treated group. On the other hand, 305 genes were detected in comparisons between the control group and the 2.5 mM phenobarbital groups (Figure 2b).

The single differentially expressed genes detected in the comparison between the control and the 0.25 mM phenobarbital-treated group was *Cricetulus griseus* early growth response 1 (Egr1), transcript variant X1, mRNA. Early growth response 1 performs the essential roles for nuclear receptor CAR to activate the human cytochrome P450 2B6 gene (Inoue & Negishi 2008; Inoue & Negishi 2009). We could obtaine the biologically significant gene from the comparison between the transcriptomes having close Kolmogorov complexity.

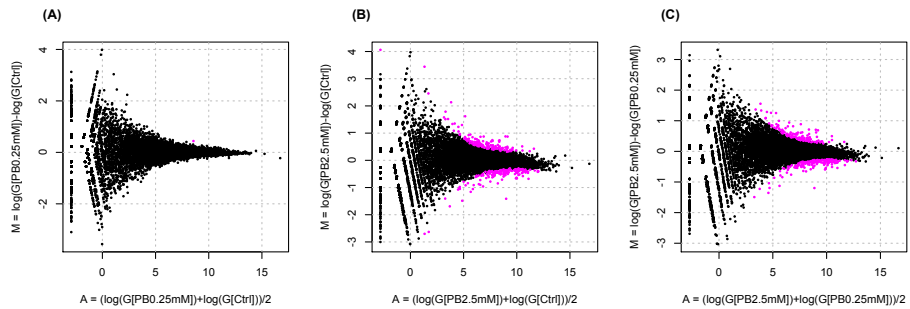

**Fig. 2. Differentially expressed gene analyses between transcriptomes.** (a) Scatter plot of differentially expressed gene analyses between Controls vs. 0.25 mM Phenobarbital experiment. (b) Scatter plot of differentially expressed gene analyses between Controls vs. 2.5 mM Phenobarbital experiment. (c) Scatter plot of differentially expressed gene analyses between 0.25 mM Phenobarbital experiment vs. 2.5 mM Phenobarbital experiment.

Previous study comparing between the amount of environmental change and the amount of transcriptome change corroborated plasticity of Shannon's information entropy of tranrciptome data and hysteretic phenomenon. The hysteretic phenomenon provides evidence of the bi-/multi-stable system. In this study, we did not examine hysteretic phenomenon although the plasticity of transcriptomes was indicated by cultivations of CHO-K1 cells in medium without phenobarbital after previous cultivation using medium containing 0.25 or 0.25 mM phenobarbital; Shannon's Information entropy were 12.01561 (0.25mM->0mM) and 11.98386 (2.5mM->0mM); Kolmogorov complexity were 0.0640789 (0.25mM->0mM) and 0.0640927 (2.5mM->0mM). Further experiments were needed.